\documentclass[10pt, conference]{IEEEtran}

\usepackage{amsfonts}
\usepackage{amssymb}
\usepackage{eurosym}
\usepackage{cite}
\usepackage{graphicx}
\usepackage{epstopdf}
\usepackage{amsmath}
\usepackage{tikz,lipsum}
\usepackage[T1]{fontenc}
\usepackage{amsthm}
\usepackage{mathrsfs}
\usepackage{color}
\usepackage{stfloats}
\usepackage{xcolor, etoolbox}
\usepackage{authblk}
\usepackage[caption=false,font=footnotesize]{subfig}

\IEEEoverridecommandlockouts

\def\BibTeX{{\rm B\kern-.05em{\sc i\kern-.025em b}\kern-.08em
    T\kern-.1667em\lower.7ex\hbox{E}\kern-.125emX}}

\begin{document}

\title{Receiver-Side Physics-Informed Residual Digital Twin for Predictive Fine Tracking in Inter-Satellite Optical Links}

\author{%
Mahtab~Dashtebozorgi,
Meysam~Ghanbari$^{1}$,
Rula~Ammuri$^{2}$,
Mazen~O.~Hasna$^{3}$,
and~Khalid~A.~Qaraqe$^{1}$\\

$^{1}$College of Science and Engineering, Hamad Bin Khalifa University, Doha, Qatar.\\
$^{2}$Professionals for Smart Technology (PST), Amman, Jordan.\\
$^{3}$Department of Electrical Engineering, Qatar University, Doha, Qatar.\\
Email: megh89467@hbku.edu.qa
}

\maketitle

\begin{abstract}
Fine tracking in inter-satellite optical links must compensate residual line-of-sight (LOS) motion despite sensor noise, vibration, model mismatch, and actuator latency. This paper develops a receiver-side physics-informed residual digital twin (PIR-Twin) that combines a nominal LOS transition, Gaussian optics, a nonlinear quadrant-photodetector observation, and extended Kalman filter synchronization. A normalized autoregressive ridge model learns the transition mismatch from independent calibration estimates, while the known delayed fine-steering-mirror correction remains separate from the physical LOS dynamics. The synchronized twin predicts the LOS at the command-actuation instant and enables proactive fine tracking. A controlled evaluation compares open-loop, reactive, nominal-predictive, and PIR-Twin operation using disjoint tuning, calibration, and test realizations. Under the nominal scenario, PIR-Twin reduces RMS pointing error by 14.1\% relative to reactive tracking and also improves actuation-time prediction accuracy. Extended robustness tests show that the proposed method retains the lowest mean pointing error over a broad actuator-delay range and under increased time-varying LOS-motion amplitudes without residual-model retraining. The results demonstrate that correcting systematic short-horizon model mismatch, rather than relying on nominal extrapolation alone, is the key mechanism enabling effective predictive fine tracking.
\end{abstract}


\section{Introduction}

Inter-satellite optical links (OISLs) support high-capacity space
connectivity through narrow, highly directional beams. Their operation
depends on accurate pointing, acquisition, and tracking (PAT), because
angular misalignment can move the beam away from the intended optical
path. On-orbit inter-satellite communication demonstrations~\cite{ref1}
and compact-terminal acquisition and tracking experiments~\cite{ref2}
illustrate the practical importance of maintaining optical alignment.
The broader PAT literature provides established coarse- and fine-steering
architectures for mobile free-space optical systems~\cite{ref3}. Once
coarse acquisition and point-ahead compensation have established the
nominal link direction, the remaining receiver-side task is to suppress
residual LOS motion within the fine-tracking range.

This residual motion can include alignment bias, slow drift, structural
vibration, and correlated jitter. A conventional feedback controller uses
an estimate of the current optical state, whereas its command may become
effective several sampling intervals later. Consequently, accurate
present-state estimation does not by itself guarantee accurate alignment
at the command-application instant. The discrepancy is especially relevant
when vibration evolves appreciably during the processing and actuation
delay. A predictor can address that discrepancy, but only if its
short-horizon dynamics adequately represent the motion to be rejected.

Quadrant photodetectors (QPDs) offer compact angular-error measurements,
and four-quadrant sensing has been investigated for tracking in mobile
optical links~\cite{ref4,Safi2021Beam}. Detector-array architectures also support joint
communication and fine beam tracking for CubeSat receivers~\cite{ref5,Safi2025CubeSat}.
State estimation is another established component: Kalman-aided
cooperative optical tracking was studied in~\cite{ref6}, and QPD
measurements have been combined with adaptive extended Kalman filtering
for inter-spacecraft laser-link acquisition~\cite{ref7}. These results
motivate an explicit optical observation model and a synchronized estimate
of the underlying LOS motion.

Predictive and learning-enhanced optical tracking has increasingly been used to mitigate sensing, modeling, and control-delay limitations. Long short-term memory prediction has been applied to pointing trajectories in composite-axis satellite laser tracking to compensate detection and control delays~\cite{ref8}. Physics-guided autoregressive identification has also been combined with disturbance-observer/model-based control and bounded learned residual compensation for LEO–GEO beam-jitter suppression~\cite{ref9}. More generally, hybrid forecasting frameworks that combine knowledge-based models with data-driven corrections have been established in other dynamical systems~\cite{ref10}.
These developments motivate a compact receiver-side formulation that combines short-horizon residual prediction with optical sensing and explicitly delayed FSM correction. Digital twins provide a suitable framework for maintaining the synchronized virtual representation required for this closed-loop prediction and control process.

Existing work addresses digital-twin-assisted optical-wireless monitoring and optimization~\cite{ref11}, optical-telescope pointing correction~\cite{ref12}, and quality-of-transmission estimation in inter-satellite all-optical networks~\cite{ref13}. Existing optical and satellite digital-twin studies primarily operate at the link, network, or telescope level, while predictive optical tracking methods generally organize estimation, prediction, and FSM control as conventional processing stages. The proposed formulation introduces a unified receiver-side architecture that synchronizes the uncompensated LOS state from nonlinear QPD measurements, learns the residual of the nominal LOS transition, and converts the resulting actuation-time forecast directly into delay-aware FSM correction. The FSM acts exclusively on the receiver optical pointing geometry, while the spacecraft/LOS state evolves independently. Learning is confined to nominal-transition mismatch using synchronized state estimates, preserving the physical model while augmenting its short-horizon predictive capability.
The principal
contributions are as follows:
\begin{itemize}
    \item A receiver-side synchronized PIR-Twin that separates physical LOS motion from delayed FSM optical correction while combining QPD sensing and EKF state estimation.

    \item A physics-informed residual predictor that preserves the nominal LOS model and learns only short-horizon transition mismatch from calibration estimates.

    \item A delay-aware FSM strategy that predicts the LOS at the command-actuation instant rather than correcting only the present state.

    \item A controlled evaluation that isolates feedback, nominal prediction, and residual learning, including robustness to larger actuator delays and stronger LOS motion without retraining.
\end{itemize}

\section{System Model}
\label{sec:system_model}

\subsection{Optical Geometry and Delayed Fine Steering}
\label{subsec:optical_geometry}

Consider an optical link between two spacecraft separated by range $L$. Coarse acquisition, nominal attitude pointing, and point-ahead compensation are assumed to have established the link, so the remaining task is receiver-side fine tracking. As illustrated in Fig.~1, the incident beam passes through the fine-steering mirror (FSM) before reaching the QPD, while the receiver-side PIR-Twin forms a synchronized digital loop using the QPD measurements and the known applied correction. The EKF updates the uncompensated LOS state, the nominal and residual models predict its future evolution, and the resulting correction is issued through the actuator-delay queue so that it becomes effective at the intended future actuation instant.
Let the uncompensated residual LOS angle and issued FSM correction be
\begin{equation}
\boldsymbol{\theta}_k
=
\begin{bmatrix}
\theta_{x,k} & \theta_{y,k}
\end{bmatrix}^{\mathrm{T}},
\qquad
\mathbf{u}_k
=
\begin{bmatrix}
u_{x,k} & u_{y,k}
\end{bmatrix}^{\mathrm{T}}.
\label{eq:los_fsm_vectors}
\end{equation}

Both quantities in \eqref{eq:los_fsm_vectors} are expressed in radians.
The command is a calibrated equivalent LOS correction, rather than a
mechanical mirror angle; any fixed optical deflection factor is included
in that calibration. For an effective delay of $d_c \geq 1$ samples, the
correction actually applied at sample $k$ is
\begin{equation}
\bar{\mathbf{u}}_k
=
\mathbf{u}_{k-d_c},
\qquad
\boldsymbol{\epsilon}_k
=
\boldsymbol{\theta}_k-\bar{\mathbf{u}}_k,
\label{eq:delayed_fsm_error}
\end{equation}
where $\boldsymbol{\epsilon}_k$ is the post-FSM pointing error. Commands
preceding the first sample are zero. This distinction is maintained in
sensing, estimation, prediction, and performance evaluation.
For a circular Gaussian beam of wavelength $\lambda$ and waist $w_0$,
the Rayleigh range and propagated $1/e^2$ radius are \cite{Ghanbari2026AIAssisted}:
\begin{equation}
z_R
=
\frac{\pi w_0^2}{\lambda},
\qquad
w_L
=
w_0
\sqrt{
1+
\left(
\frac{L}{z_R}
\right)^2
}.
\label{eq:gaussian_beam_radius}
\end{equation}

\begin{figure}[t]
    \centering
    \includegraphics[width=\columnwidth]{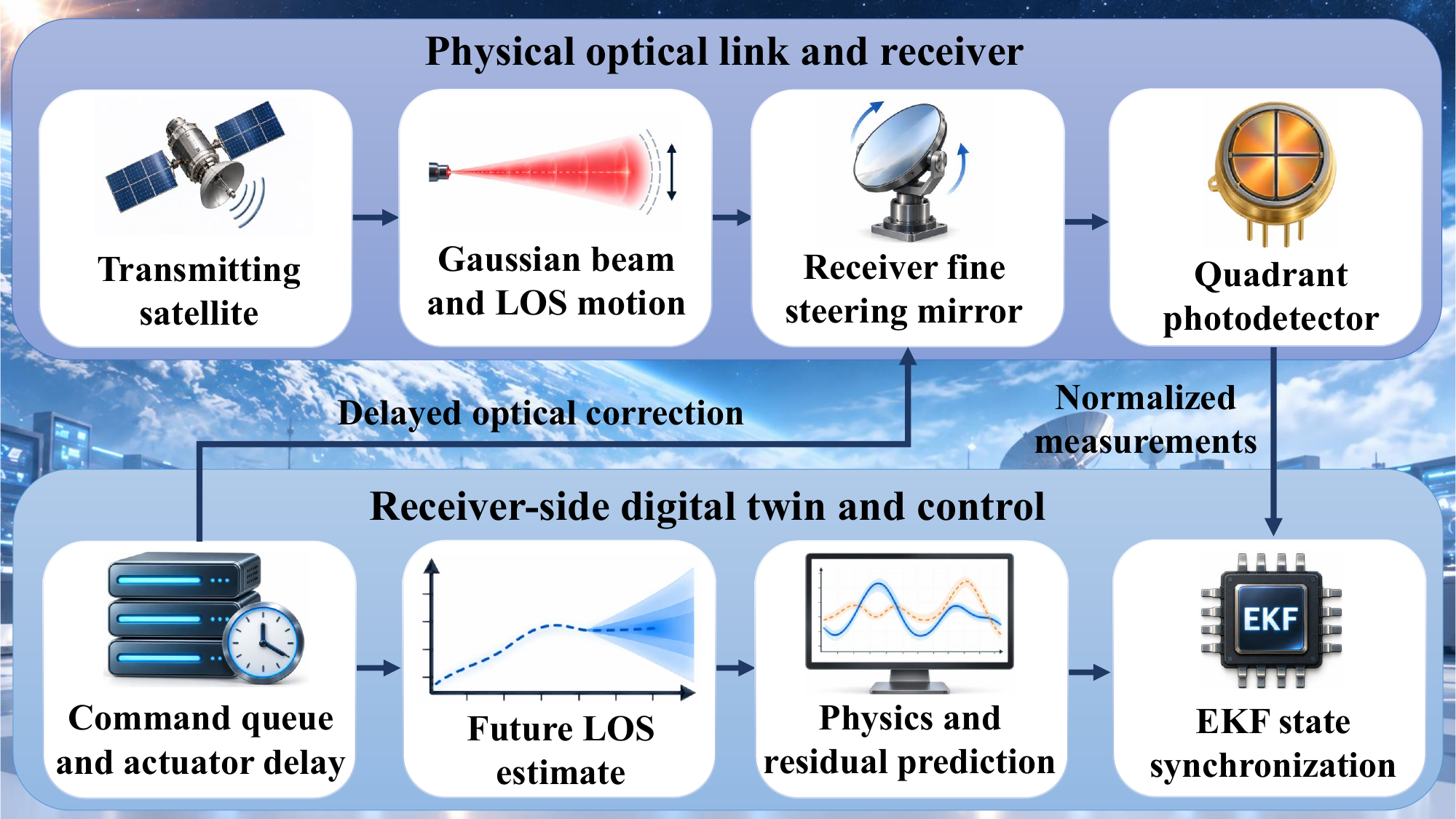}
\caption{Receiver-side PIR-Twin architecture for synchronized LOS estimation, prediction, and delayed FSM correction.}
    \label{fig:fig1}
\end{figure}

The centered fraction collected by a circular aperture of radius $a_r$
and the aligned optical reference are \cite{ref4}:
\begin{equation}
A_0
=
1-
\exp\left(
-\frac{2a_r^2}{w_L^2}
\right),
\qquad
P_{r,0}
=
P_t \eta_t \eta_r A_0,
\label{eq:aperture_collection}
\end{equation}
with transmitted power $P_t$ and fixed terminal efficiencies
$\eta_t$ and $\eta_r$. These expressions provide the optical scale used
by the simulator.
The reported angular-quality metric uses the equivalent Gaussian penalty 
\begin{equation}
\eta_{p,k}
=
\exp\left(
-\frac{2L^2
\left\lVert
\boldsymbol{\epsilon}_k
\right\rVert^2}
{w_L^2}
\right),
\qquad
P_{\mathrm{proxy},k}
=
P_{r,0}\eta_{p,k}.
\label{eq:gaussian_pointing_proxy}
\end{equation}

The quantity $P_{\mathrm{proxy},k}$ provides an equivalent Gaussian angular-quality metric that maps the residual post-FSM pointing error to the corresponding normalized optical-alignment quality.
The tracking controller operates directly on the synchronized angular state, while the metric in \eqref{eq:gaussian_pointing_proxy} provides a complementary optical-quality characterization of the resulting residual alignment.
This formulation provides a consistent mapping between residual angular tracking accuracy and the corresponding equivalent optical-alignment quality.

\subsection{Normalized QPD Observation}
\label{subsec:qpd_observation}

The post-FSM angle determines the focal-plane beam center, \cite{Ghanbari2026AIAssisted}:
\begin{equation}
x_{b,k}
=
f\epsilon_{x,k},
\qquad
y_{b,k}
=
f\epsilon_{y,k},
\label{eq:qpd_beam_center}
\end{equation}
where $f$ is the effective focal length. With focal-spot standard
deviation $\sigma_f$, a unit-normalized Gaussian intensity is
\begin{equation}
I_k(\xi,\nu)
=
\frac{1}{2\pi\sigma_f^2}
\exp
\left[
-\frac{
(\xi-x_{b,k})^2
+
(\nu-y_{b,k})^2
}{
2\sigma_f^2
}
\right].
\label{eq:qpd_gaussian_intensity}
\end{equation}

For quadrant region $\mathcal{Q}_i$, let
\begin{equation}
q_{i,k}
=
\iint_{\mathcal{Q}_i}
I_k(\xi,\nu)\, d\xi\, d\nu,
\qquad
\sum_{i=1}^{4} q_{i,k}=1.
\label{eq:qpd_quadrant_fraction}
\end{equation}

The quadrants are numbered counterclockwise from the upper right. 
A canonical four-quadrant QPD model with sufficiently large active area and negligible inter-quadrant gap is adopted for fine-tracking characterization.
Its
normalized differential outputs are therefore \cite{ref5}:
\begin{equation}
\begin{aligned}
S_{x,k}
&=
\frac{
q_{1,k}+q_{4,k}-q_{2,k}-q_{3,k}
}{
\sum_i q_{i,k}
},
\\[3pt]
S_{y,k}
&=
\frac{
q_{1,k}+q_{2,k}-q_{3,k}-q_{4,k}
}{
\sum_i q_{i,k}
}.
\end{aligned}
\label{eq:qpd_differential_outputs}
\end{equation}

Gaussian separability reduces the quadrant integrals in
\eqref{eq:qpd_quadrant_fraction} to the nonlinear measurement model
\begin{equation}
\mathbf{y}_k
=
h(\boldsymbol{\epsilon}_k)
+
\mathbf{v}_k,
\qquad
h(\boldsymbol{\epsilon})
=
\begin{bmatrix}
\operatorname{erf}(\alpha \epsilon_x)
\\
\operatorname{erf}(\alpha \epsilon_y)
\end{bmatrix},
\label{eq:qpd_measurement_model}
\end{equation}
where 
$ 
\alpha
=
\frac{f}{\sqrt{2}\sigma_f}$
and
$
\mathbf{v}_k
\sim
\mathcal{N}(\mathbf{0},\mathbf{R})
$
is normalized additive measurement noise. The error-function response and
its analytic derivative are used directly in the implementation. Its
sensitivity decreases away from the detector center, although the
evaluated fine-pointing angles lie close to the linear operating region.
The normalized QPD measurement uncertainty is represented by a fixed covariance \(R\), providing a controlled sensing condition for evaluating state synchronization and predictive tracking.

\subsection{LOS State and Nominal Dynamics}
\label{subsec:los_dynamics}

The state ordering is
\begin{equation}
\mathbf{x}_k
=
\begin{bmatrix}
\theta_{x,k} &
\omega_{x,k} &
\theta_{y,k} &
\omega_{y,k}
\end{bmatrix}^{\mathrm{T}},
\qquad
\boldsymbol{\theta}_k
=
\mathbf{C}\mathbf{x}_k,
\label{eq:state_definition}
\end{equation}
where $\omega_{x,k}$ and $\omega_{y,k}$ are angular rates. The nominal
constant-angular-rate transition and angle selector are
\begin{equation}
\mathbf{A}
=
\begin{bmatrix}
1 & T_s & 0 & 0 \\
0 & 1   & 0 & 0 \\
0 & 0   & 1 & T_s \\
0 & 0   & 0 & 1
\end{bmatrix},
\qquad
\mathbf{C}
=
\begin{bmatrix}
1 & 0 & 0 & 0 \\
0 & 0 & 1 & 0
\end{bmatrix}.
\label{eq:nominal_transition}
\end{equation}

The exact discrepancy relative to the nominal model in
\eqref{eq:nominal_transition} is written as
\begin{equation}
\mathbf{x}_{k+1}
=
\mathbf{A}\mathbf{x}_k
+
\mathbf{r}_k,
\qquad
\mathbf{r}_k
=
\mathbf{x}_{k+1}
-
\mathbf{A}\mathbf{x}_k.
\label{eq:transition_residual}
\end{equation}

The residual in \eqref{eq:transition_residual} includes the structured
short-horizon motion omitted by constant-rate propagation. 
The physical LOS trajectory is generated independently from the nominal estimator dynamics using a richer bounded motion model, thereby enabling genuine transition-mismatch evaluation.
Combining the state relation in \eqref{eq:state_definition}, the delayed
fine-steering relation in \eqref{eq:delayed_fsm_error}, and the nonlinear
QPD observation in \eqref{eq:qpd_measurement_model} gives
\begin{equation}
\mathbf{y}_k
=
h
\left(
\mathbf{C}\mathbf{x}_k
-
\mathbf{u}_{k-d_c}
\right)
+
\mathbf{v}_k.
\label{eq:combined_measurement}
\end{equation}

The applied correction is known to the receiver. It enters
\eqref{eq:combined_measurement} as an optical offset and does not enter
\eqref{eq:transition_residual} as a spacecraft acceleration. At the loop
level, the FSM is represented by calibrated angular authority and a fixed
effective delay; the commanded variable is the absolute correction angle.

\section{Receiver-Side PIR-Twin and Predictive Control}
\label{sec:pir_twin}

\subsection{Twin Representation and Synchronization}
\label{subsec:twin_representation}

The residual regressor is one component of the twin. At sample $k$, the
receiver maintains the synchronized representation
\begin{equation}
\mathcal{T}_k
=
\left\{
\hat{\mathbf{x}}_{k|k},
\mathbf{P}_{k|k},
\mathbf{A},
\mathbf{C},
h,
\mathcal{M}_r,
\mathcal{U}_k,
\Xi
\right\},
\label{eq:twin_representation}
\end{equation}
where $\mathcal{M}_r$ denotes the residual model, $\mathcal{U}_k$ the
command history required to implement delay, and $\Xi$ the optical and
actuator parameters. Thus, the twin contains both the synchronized
physical state and the models needed to interpret measurements and
forecast optical alignment.
Each sample begins by applying the oldest queued correction. The QPD then
measures the resulting optical error, and an extended Kalman filter (EKF)
updates the uncompensated LOS estimate using that known correction. The
posterior state initializes the predictor, which advances to the actuation
instant of a newly issued command. That command is appended to the queue.
This order provides a continuous measurement-to-twin synchronization path
and a prediction-to-actuator control path.
The residual model is not inserted into the EKF time update. Keeping the
same nominal estimator for the reactive and predictive controllers
isolates the contribution of residual-enhanced forecasting.
For clarity, we write
$
\hat{\mathbf{x}}_k^{-}
\equiv
\hat{\mathbf{x}}_{k|k-1}$,
$
\hat{\mathbf{x}}_k^{+}
\equiv
\hat{\mathbf{x}}_{k|k},
$
with analogous notation for covariance. The EKF time update is \cite{Grewal2015Kalman}:
\begin{equation}
\hat{\mathbf{x}}_k^{-}
=
\mathbf{A}\hat{\mathbf{x}}_{k-1}^{+},
\qquad
\mathbf{P}_k^{-}
=
\mathbf{A}\mathbf{P}_{k-1}^{+}\mathbf{A}^{\mathrm{T}}
+
\mathbf{Q}_E.
\label{eq:ekf_time_update}
\end{equation}

At the first sample, the prescribed initial state and covariance are used
directly. The assumed process covariance is \cite{Grewal2015Kalman}:
\begin{equation}
\mathbf{Q}_E
=
q_a
\operatorname{diag}
\left(
\mathbf{g}\mathbf{g}^{\mathrm{T}},
\mathbf{g}\mathbf{g}^{\mathrm{T}}
\right),
\qquad
\mathbf{g}
=
\begin{bmatrix}
T_s^2/2 \\
T_s
\end{bmatrix}.
\label{eq:ekf_process_covariance}
\end{equation}

This is the discrete covariance associated with a per-interval
angular-acceleration variance scale $q_a$, in
$\mathrm{rad}^2\mathrm{s}^{-4}$. 
The predicted pointing error, QPD output, and measurement Jacobian are \cite{Grewal2015Kalman}:
\begin{equation}
\hat{\boldsymbol{\epsilon}}_k^{-}
=
\mathbf{C}\hat{\mathbf{x}}_k^{-}
-
\bar{\mathbf{u}}_k,
\qquad
\hat{\mathbf{y}}_k
=
h\!\left(
\hat{\boldsymbol{\epsilon}}_k^{-}
\right),
\label{eq:predicted_pointing_qpd}
\end{equation}

\begin{equation}
\mathbf{H}_k
=
\operatorname{diag}
\left(
\beta_{x,k},
\beta_{y,k}
\right)
\mathbf{C},
\qquad
\beta_{i,k}
=
\frac{2\alpha}{\sqrt{\pi}}
\exp
\left[
-\alpha^2
\left(
\hat{\epsilon}_{i,k}^{-}
\right)^2
\right],
\label{eq:qpd_measurement_jacobian}
\end{equation}
for $i\in\{x,y\}$. Defining the innovation and its covariance as \cite{Grewal2015Kalman}:
\begin{equation}
\boldsymbol{\nu}_k
=
\mathbf{y}_k
-
\hat{\mathbf{y}}_k,
\qquad
\mathbf{S}_k
=
\mathbf{H}_k
\mathbf{P}_k^{-}
\mathbf{H}_k^{\mathrm{T}}
+
\mathbf{R},
\label{eq:innovation_covariance}
\end{equation}
the gain and posterior state are \cite{Grewal2015Kalman}:
\begin{equation}
\mathbf{K}_k
=
\mathbf{P}_k^{-}
\mathbf{H}_k^{\mathrm{T}}
\mathbf{S}_k^{-1},
\qquad
\hat{\mathbf{x}}_k^{+}
=
\hat{\mathbf{x}}_k^{-}
+
\mathbf{K}_k
\boldsymbol{\nu}_k.
\label{eq:kalman_gain_posterior}
\end{equation}

The covariance uses the Joseph form, \cite{Grewal2015Kalman}:
\begin{equation}
\begin{aligned}
\mathbf{P}_k^{+}
={}&
\left(
\mathbf{I}
-
\mathbf{K}_k\mathbf{H}_k
\right)
\mathbf{P}_k^{-}
\left(
\mathbf{I}
-
\mathbf{K}_k\mathbf{H}_k
\right)^{\mathrm{T}}
\\
&+
\mathbf{K}_k
\mathbf{R}
\mathbf{K}_k^{\mathrm{T}},
\end{aligned}
\label{eq:joseph_covariance}
\end{equation}
followed by numerical symmetrization. Angular position is sensed directly,
while the rate components are inferred through the state transition and
repeated measurements. Accounting for $\bar{\mathbf{u}}_k$ allows the EKF
to track uncompensated LOS motion even though the detector observes only
the post-FSM error.

\subsection{Physics-Informed Residual Calibration}
\label{subsec:residual_calibration}

The nominal transition provides a local motion model, while vibration and
drift produce the residual in \eqref{eq:transition_residual}. Learning is
confined to this transition mismatch. The calibration data available to
the receiver are synchronized EKF posterior estimates rather than the true
simulated state. We denote these estimates by
$
\bar{\mathbf{x}}_k
=
\hat{\mathbf{x}}_{k|k}^{\mathrm{cal}}
$
and define
\begin{equation}
\bar{\mathbf{r}}_k
=
\bar{\mathbf{x}}_{k+1}
-
\mathbf{A}\bar{\mathbf{x}}_k.
\label{eq:calibration_residual}
\end{equation}

Unlike $r_k$ in \eqref{eq:transition_residual}, this observable target is constructed directly from synchronized EKF posterior states and therefore represents the receiver-accessible posterior transition mismatch used for residual calibration.
The calibration filter uses
the same nominal dynamics as the online estimator.
Let
$
\mathbf{D}_x
=
\operatorname{diag}
\left(
s_1,s_2,s_3,s_4
\right)
$
contain positive state scales. Current and preceding posterior estimates
form the normalized autoregressive feature,
\begin{equation}
\boldsymbol{\phi}_k
=
\begin{bmatrix}
\mathbf{D}_x^{-1}\bar{\mathbf{x}}_k
\\
\mathbf{D}_x^{-1}\bar{\mathbf{x}}_{k-1}
\\
1
\end{bmatrix}
\in
\mathbb{R}^{9},
\qquad
\mathbf{z}_k
=
\mathbf{D}_x^{-1}\bar{\mathbf{r}}_k.
\label{eq:normalized_residual_features}
\end{equation}

The previous state supplies short-term temporal information, while scaling
prevents the different magnitudes and units of angles and rates from
dominating the fit. The intercept accommodates a constant component of the
residual. 
After discarding an initial synchronization interval within each
calibration trace, the valid features and targets are pooled as columns of
$\boldsymbol{\Phi}$ and $\mathbf{Z}$. The ridge objective is
\begin{equation}
\min_{\mathbf{W}}
\;
\left\|
\mathbf{Z}
-
\mathbf{W}\boldsymbol{\Phi}
\right\|_{F}^{2}
+
\lambda_r
\left\|
\mathbf{W}\boldsymbol{\Gamma}^{1/2}
\right\|_{F}^{2}.
\label{eq:ridge_objective}
\end{equation}

Here
$
\boldsymbol{\Gamma}
=
\operatorname{diag}(1,\ldots,1,0)
$
leaves the intercept unregularized. The implemented solution, with no
normalization by the sample count, is
\begin{equation}
\mathbf{W}
=
\mathbf{Z}\boldsymbol{\Phi}^{\mathrm{T}}
\left(
\boldsymbol{\Phi}\boldsymbol{\Phi}^{\mathrm{T}}
+
\lambda_r \boldsymbol{\Gamma}
\right)^{-1}.
\label{eq:ridge_solution}
\end{equation}

For a feature vector $\boldsymbol{\phi}$, the physical residual prediction
and its bounded form are
\begin{equation}
\hat{\mathbf{r}}
=
\mathbf{D}_x \mathbf{W}\boldsymbol{\phi},
\qquad
\hat{\mathbf{r}}^{\,c}
=
\operatorname{clip}
\left(
\hat{\mathbf{r}},
-\mathbf{r}_{\max},
\mathbf{r}_{\max}
\right).
\label{eq:bounded_residual_prediction}
\end{equation}

Clipping acts separately on angle and rate increments. It bounds
extrapolated corrections without removing the nominal transition. The
resulting predictor is an explicit physical model augmented by a small
learned residual map; its weights are fixed before controller testing.

\subsection{Prediction and Pending Commands}
\label{subsec:prediction_pending_commands}

At sample $k$, the prediction horizon equals the command delay, $d_c$.
The physics-only forecast is initialized by the current posterior and
propagated as
\begin{equation}
\tilde{\mathbf{x}}_{0|k}^{\mathrm{PHY}}
=
\hat{\mathbf{x}}_{k|k},
\qquad
\tilde{\mathbf{x}}_{j+1|k}^{\mathrm{PHY}}
=
\mathbf{A}
\tilde{\mathbf{x}}_{j|k}^{\mathrm{PHY}},
\label{eq:physics_only_prediction}
\end{equation}
for $j=0,\ldots,d_c-1$. It gives the baseline actuation-time angle
$\mathbf{C}\tilde{\mathbf{x}}_{d_c|k}^{\mathrm{PHY}}$.
For PIR-Twin, set
$
\tilde{\mathbf{x}}_{0|k}
=
\hat{\mathbf{x}}_{k|k}$,
$
\tilde{\mathbf{x}}_{-1|k}
=
\hat{\mathbf{x}}_{k-1|k-1}$.
At the first sample, the current posterior is duplicated to initialize the
history. Each forward step uses the recursively predicted current and
preceding states:
\begin{equation}
\tilde{\boldsymbol{\phi}}_{j|k}
=
\begin{bmatrix}
\mathbf{D}_x^{-1}\tilde{\mathbf{x}}_{j|k}
\\
\mathbf{D}_x^{-1}\tilde{\mathbf{x}}_{j-1|k}
\\
1
\end{bmatrix},
\label{eq:recursive_feature}
\end{equation}

\begin{equation}
\hat{\mathbf{r}}_{j|k}^{\,c}
=
\operatorname{clip}
\left(
\mathbf{D}_x
\mathbf{W}
\tilde{\boldsymbol{\phi}}_{j|k},
-\mathbf{r}_{\max},
\mathbf{r}_{\max}
\right),
\label{eq:recursive_residual}
\end{equation}

\begin{equation}
\tilde{\mathbf{x}}_{j+1|k}
=
\mathbf{A}
\tilde{\mathbf{x}}_{j|k}
+
\hat{\mathbf{r}}_{j|k}^{\,c}.
\label{eq:pir_twin_rollout}
\end{equation}

After each step, the newly predicted state becomes the current state and
the preceding current state becomes its history. This is a recursive
state-transition predictor, rather than a separate regressor fitted for
each actuator delay. Changing $d_c$ changes the rollout length while
retaining the same residual model.
The forecasts in \eqref{eq:physics_only_prediction} and
\eqref{eq:pir_twin_rollout} describe uncompensated LOS motion. The commands
already issued before sample $k$ determine optical correction within the
pending delay window. Specifically, for $1 \leq j < d_c$,
\begin{equation}
\bar{\mathbf{u}}_{k+j|k}
=
\mathbf{u}_{k+j-d_c},
\qquad
\tilde{\boldsymbol{\epsilon}}_{k+j|k}
=
\mathbf{C}
\tilde{\mathbf{x}}_{j|k}
-
\mathbf{u}_{k+j-d_c}.
\label{eq:pending_command_error}
\end{equation}

At the new command's actuation instant,
\begin{equation}
\tilde{\boldsymbol{\epsilon}}_{k+d_c|k}
=
\mathbf{C}
\tilde{\mathbf{x}}_{d_c|k}
-
\mathbf{u}_k.
\label{eq:actuation_instant_error}
\end{equation}

Thus, pending corrections are known optical offsets; they do not need to be
inserted into the independent LOS state transition. The implementation
forecasts the actuation-time LOS directly and realizes the intermediate
corrections through a first-in, first-out (FIFO) command queue. No
optimization over unknown future commands is required.

\subsection{Controller Definitions}
\label{subsec:controller_definitions}

The controller comparison distinguishes feedback, nominal prediction, and
residual correction. All commands use the same gain $\kappa$ and per-axis
authority $u_{\max}$. With $\operatorname{sat}_{u_{\max}}$ denoting
componentwise saturation, the four modes are
\begin{equation}
\begin{aligned}
\mathbf{u}_k^{\mathrm{OL}}
&=
\mathbf{0},
\\
\mathbf{u}_k^{\mathrm{R}}
&=
\operatorname{sat}_{u_{\max}}
\left(
\kappa \mathbf{C}\hat{\mathbf{x}}_{k|k}
\right),
\\
\mathbf{u}_k^{\mathrm{PHY}}
&=
\operatorname{sat}_{u_{\max}}
\left(
\kappa
\mathbf{C}
\mathbf{A}^{d_c}
\hat{\mathbf{x}}_{k|k}
\right),
\\
\mathbf{u}_k^{\mathrm{PIR}}
&=
\operatorname{sat}_{u_{\max}}
\left(
\kappa
\mathbf{C}
\tilde{\mathbf{x}}_{d_c|k}
\right).
\end{aligned}
\label{eq:controller_modes}
\end{equation}

Open loop exposes the uncompensated trajectory. Reactive tracking commands
the present estimated angle, so its implicit forecast assumes that angle
persists until actuation. Nominal prediction extrapolates the current state
through the constant-rate model. PIR-Twin uses the same horizon, with the
learned transition correction added at every step.
Saturation acts elementwise as
$
\left[
\operatorname{sat}_{u_{\max}}(\mathbf{a})
\right]_i
=
\min
\left\{
u_{\max},
\max
\left[
-u_{\max},
a_i
\right]
\right\}$.
The calibrated gain is $\kappa=1$ throughout the evaluation. Each command
specifies the absolute correction that replaces the previous setting when
it reaches the FIFO output.

\section{Numerical Results}
\label{sec:numerical_results}

\subsection{Physical Scenario and Data Separation}
\label{subsec:physical_scenario}

The simulation represents a 1000-km link at $\lambda=1550~\mathrm{nm}$
with $w_0=30~\mathrm{mm}$, giving a $16.45$-$\mu\mathrm{rad}$ Gaussian
$1/e^2$ half-angle. The receive-aperture diameter is $100~\mathrm{mm}$.
The optical reference uses $P_t=100~\mathrm{mW}$ and efficiencies
$\eta_t=0.70$ and $\eta_r=0.65$. Fine tracking operates at $1~\mathrm{kHz}$
with a nominal two-sample ($2$-ms) delay and a
$\pm 50$-$\mu\mathrm{rad}$ correction limit per axis. The QPD has
$f=80~\mathrm{mm}$, $\sigma_f=0.18~\mathrm{mm}$, and normalized noise
covariance
$
\mathbf{R}
=
(2\times10^{-4})^2\mathbf{I}$.

These parameters define a representative long-range OISL fine-tracking regime.
The physical truth is generated independently of the nominal constant-rate
model. Each axis contains a static bias, slow drift, two vibration
components, and three smaller random low-frequency tones. The static bias
is
$
\begin{bmatrix}
20 & -15
\end{bmatrix}^{\mathrm{T}}
~\mu\mathrm{rad}$;
the drift amplitude is $5~\mu\mathrm{rad}$ at $0.70$ and $0.55~\mathrm{Hz}$
along the respective axes. The $35$- and $80$-Hz vibration amplitudes are
$3$ and $1~\mu\mathrm{rad}$. Additional tones have frequencies uniformly
selected between $2$ and $12~\mathrm{Hz}$ and nominal amplitude
$0.25~\mu\mathrm{rad}$, multiplied by a uniform factor between $0.7$ and
$1.3$. Phases and the small-tone realizations change with the seed. Angular
rates are the exact derivatives of the generated continuous trajectories.
No FSM command or independent white state-process noise is added to this
physical truth.
The EKF starts from a zero state estimate with covariance
$
\operatorname{diag}
\left(
5\times10^{-9},
5\times10^{-5},
5\times10^{-9},
5\times10^{-5}
\right)
$
in the corresponding mixed state units. A separate $2$-s tuning trace,
seed~9, selects the covariance scale using angle-estimation
root-mean-square error (RMSE), rather than controller performance. The nine
candidate scales are
$
0.001,\;
0.003,\;
0.01,\;
0.03,\;
0.1,\;
0.3,\;
1,\;
3,\;
10.
$

The selected value is
$
q_a=0.1,
$
yielding $0.66~\mu\mathrm{rad}$ tuning RMSE. This covariance is then fixed
for calibration, controller testing, and all delays.
Four independent 5-s calibration traces use seeds 1, 3, 5, and 7.
Calibration is open loop, so the applied optical correction is zero. The
first $0.5$~s of each trace is excluded from residual fitting; valid
adjacent transitions are formed within each trace and then pooled. This
gives 4500 samples per trace and 18\,000 samples in total. The state
normalization and residual bounds are
\begin{equation}
\mathbf{D}_x
=
\operatorname{diag}
\left(
50~\mu\mathrm{rad},
5~\mathrm{mrad/s},
50~\mu\mathrm{rad},
5~\mathrm{mrad/s}
\right),
\label{eq:state_scaling}
\end{equation}

\begin{equation}
\mathbf{r}_{\max}
=
\begin{bmatrix}
2~\mu\mathrm{rad} &
2~\mathrm{mrad/s} &
2~\mu\mathrm{rad} &
2~\mathrm{mrad/s}
\end{bmatrix}^{\mathrm{T}}.
\label{eq:residual_bounds}
\end{equation}

The ridge coefficient is
$
\lambda_r = 2\times10^{-4}$.
The fitted weights remain fixed throughout the evaluation.
The primary evaluation uses ten previously unseen 2-s test traces, with tuning, calibration, and testing performed on mutually disjoint realizations. All controller modes share the same physical LOS trajectory, QPD-noise realization, estimator, actuator limits, and control gain for each test case. Robustness is further examined by scaling the time-varying drift, vibration, and random-tone amplitudes by $s \in \{1, 1.5, 2\}$, while the remaining system parameters, EKF settings, and trained residual model are kept fixed without retraining or retuning.

\begin{table*}[t]
\centering
\caption{Primary Controller Comparison Over Ten Independent Test Realizations at 2-ms FSM Delay}
\label{tab:primary_controller_comparison}
\renewcommand{\arraystretch}{1.05}
\setlength{\tabcolsep}{8pt}
\begin{tabular}{lcccc}
\hline
Method
&
Pointing RMS [$\mu$rad]
&
Prediction RMSE [$\mu$rad]
&
Angular Penalty [dB]
&
Command RMS [$\mu$rad]
\\
\hline
Open loop
&
$25.7 \pm 0.6$
&
---
&
$16.73$
&
$0.0$
\\
Reactive
&
$2.12 \pm 0.08$
&
$1.95 \pm 0.01$
&
$0.125$
&
$25.7$
\\
Physics predictive
&
$2.11 \pm 0.08$
&
$1.93 \pm 0.01$
&
$0.123$
&
$25.9$
\\
PIR-Twin predictive
&
$\mathbf{1.82 \pm 0.11}$
&
$\mathbf{1.62 \pm 0.05}$
&
$\mathbf{0.088}$
&
$\mathbf{25.7}$
\\
\hline
\end{tabular}

\vspace{2pt}
\begin{minipage}{0.97\textwidth}
\footnotesize
Entries are means across ten test seeds; $\pm$ indicates sample standard
deviation.
\end{minipage}
\end{table*}

\begin{figure*}[t]
    \centering

    \subfloat[]{%
        \includegraphics[width=0.32\textwidth]{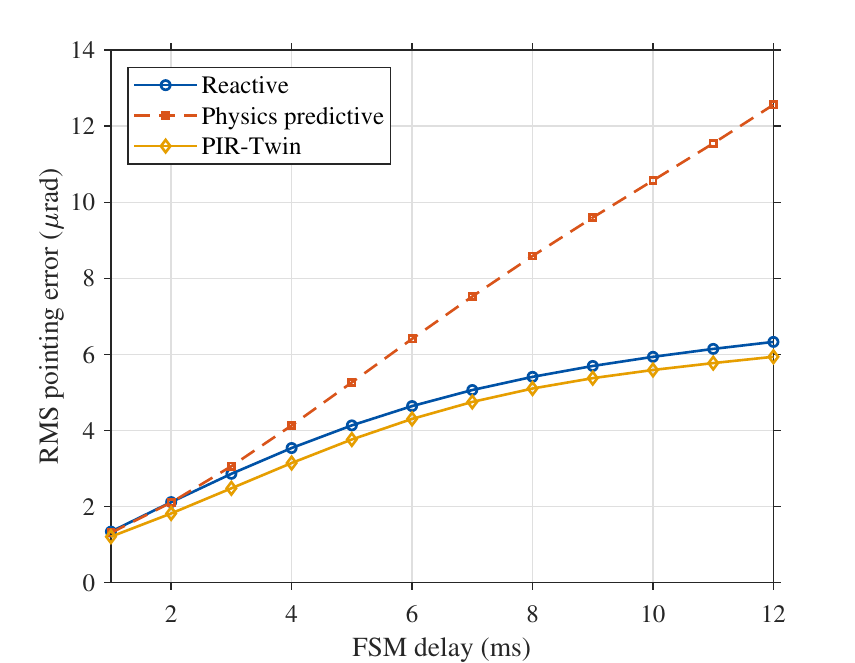}
        \label{fig:robust_nominal}
    }
    \hfill
    \subfloat[]{%
        \includegraphics[width=0.32\textwidth]{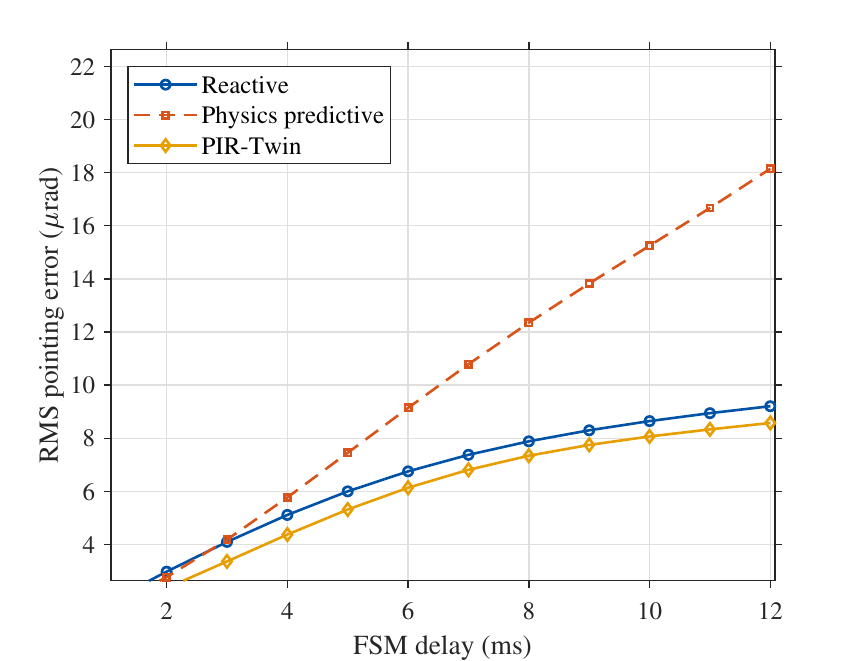}
        \label{fig:robust_15x}
    }
    \hfill
    \subfloat[]{%
        \includegraphics[width=0.32\textwidth]{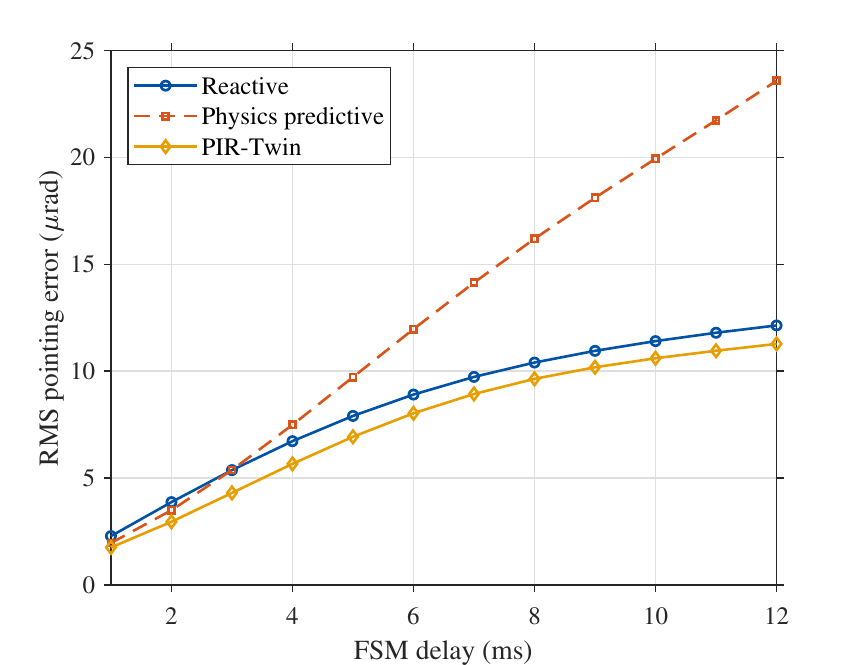}
        \label{fig:robust_2x}
    }

    \caption{Mean RMS post-FSM pointing error versus FSM delay for (a) nominal, (b) $1.5\times$, and (c) $2\times$ time-varying LOS-motion amplitudes, while the remaining system parameters and the trained residual model remain fixed.}
    \label{fig:robustness_delay}
\end{figure*}

\subsection{Metrics and Their Relation to Delay}
\label{subsec:metrics_delay}

For each trace of $N=2001$ samples, the two-axis pointing RMS and
applied-command RMS are
\begin{equation}
E_{\mathrm{pt}}
=
\sqrt{
\frac{1}{N}
\sum_{k=1}^{N}
\left\|
\boldsymbol{\epsilon}_k
\right\|^2
},
\qquad
U_{\mathrm{rms}}
=
\sqrt{
\frac{1}{N}
\sum_{k=1}^{N}
\left\|
\bar{\mathbf{u}}_k
\right\|^2
}.
\label{eq:pointing_command_rms}
\end{equation}

Pointing RMS includes the initial samples before any command becomes
effective. 
Command RMS quantifies the magnitude of the applied equivalent LOS correction, including bias compensation, and enables direct comparison of actuator utilization across the controlled methods.
The equivalent angular penalty is
\begin{equation}
L_{p,\mathrm{dB}}
=
-10
\log_{10}
\left(
\frac{1}{N}
\sum_{k=1}^{N}
\eta_{p,k}
\right).
\label{eq:angular_loss_proxy}
\end{equation}

This is the loss of mean normalized efficiency, rather than the mean
instantaneous loss in decibels. The per-trace dB values are averaged
across seeds.
Let
$\hat{\boldsymbol{\theta}}_{k+d_c|k}$
be the angle used to generate the command: the present posterior angle for
reactive control, or the appropriate future prediction for the predictive
methods. Its actuation-time RMSE is
\begin{equation}
E_{\mathrm{pred}}
=
\sqrt{
\frac{1}{N-d_c}
\sum_{k=1}^{N-d_c}
\left\|
\hat{\boldsymbol{\theta}}_{k+d_c|k}
-
\mathbf{C}\mathbf{x}_{k+d_c}
\right\|^2
}.
\label{eq:actuation_prediction_rmse}
\end{equation}

Only issue times with available future truth are included. With unity gain
and inactive command saturation, this forecast determines the delayed
tracking error exactly:
\begin{equation}
\boldsymbol{\epsilon}_{k+d_c}
=
\mathbf{C}\mathbf{x}_{k+d_c}
-
\hat{\boldsymbol{\theta}}_{k+d_c|k}.
\label{eq:forecast_tracking_relation}
\end{equation}

Consequently, for each test trace and controlled mode,
\begin{equation}
N E_{\mathrm{pt}}^2
=
\sum_{k=1}^{d_c}
\left\|
\mathbf{C}\mathbf{x}_k
\right\|^2
+
(N-d_c)E_{\mathrm{pred}}^2.
\label{eq:rms_relation}
\end{equation}

The distinction between the two RMS measures is therefore startup
accounting. Prediction RMSE directly characterizes the error inherited by
the delayed controller, while complete-record pointing RMS also includes
initial alignment before the FIFO fills. All tabulated means are averages
of per-seed metrics, not a single RMS pooled over all samples; reported
standard deviations describe variation across seeds.

\subsection{Primary Controller Comparison}
\label{subsec:primary_comparison}

Table~\ref{tab:primary_controller_comparison} first establishes the benefit
of closing the fine-tracking loop: reactive EKF/FSM operation reduces mean
pointing RMS from $25.7$ to $2.12~\mu\mathrm{rad}$, a $91.8\%$ reduction.
Reactive EKF/FSM operation therefore establishes a strong closed-loop tracking baseline, providing a demanding reference against which the additional predictive capability of PIR-Twin is evaluated.

At the nominal 2-ms delay, constant-rate prediction reaches
$2.11~\mu\mathrm{rad}$, only $0.5\%$ below reactive tracking. 
The near-identical reactive and physics-predictive results show that nominal constant-rate extrapolation alone cannot fully exploit the short-horizon structure of the LOS dynamics.
PIR-Twin reduces the mean to $1.82~\mu\mathrm{rad}$,
corresponding to $14.1\%$ and $13.6\%$ improvements over the reactive and
physics-predictive baselines. The actuation-time prediction RMSE decreases
from $1.95$ and $1.93~\mu\mathrm{rad}$ to $1.62~\mu\mathrm{rad}$. Through
\eqref{eq:forecast_tracking_relation}, this lower forecast error translates
directly into improved alignment once commands become active.

The equivalent angular-quality metric follows the same trend, decreasing from 0.125 dB for reactive tracking to 0.088 dB for PIR-Twin and consistently reflecting the improved residual alignment.
 Applied-command RMS is approximately
$25.7$--$25.9~\mu\mathrm{rad}$ for all controlled methods, and none reaches
the $\pm 50$-$\mu\mathrm{rad}$ per-axis limit. The improvement therefore
occurs with similar correction amplitudes and identical actuator authority.
RMS pointing error is adopted as the principal alignment metric because it characterizes sustained closed-loop tracking performance over the complete realization.

\subsection{Robustness to Actuator Delay and LOS-Motion Amplitude}
\label{subsec:delay_sensitivity}
Fig.~2 evaluates robustness to actuator delay and increased time-varying LOS-motion amplitude.
In Fig.~2(a), PIR-Twin achieves the lowest mean RMS pointing error throughout the 1--12-ms delay range. At the nominal 2-ms delay, the reactive, physics-predictive, and PIR-Twin errors are 2.12, 2.11, and 1.82~$\mu$rad, respectively. As the delay increases, constant-rate prediction degrades rapidly because model mismatch accumulates over the longer prediction horizon; at 12~ms, its error reaches 12.56~$\mu$rad, compared with 6.33 and 5.94~$\mu$rad for reactive and PIR-Twin tracking.
Figs.~2(b) and 2(c) show that stronger LOS motion further increases the benefit of residual correction. At 2~ms, the PIR-Twin improvement over reactive tracking rises from 14.1\% in Fig.~2(a) to 20.7\% and 23.7\% in Figs.~2(b) and 2(c), respectively. PIR-Twin remains below the reactive baseline for every tested delay and amplitude combination, whereas nominal prediction becomes increasingly unreliable as the horizon grows. The most severe cases begin to encounter the finite FSM authority, with a maximum mean saturation fraction of approximately 6.5\%. 
These results demonstrate that the fixed PIR-Twin residual model preserves its predictive advantage across substantial variations in actuator delay and LOS-motion amplitude without retraining or controller retuning.

\section{Conclusion}
\label{sec:conclusion}

A receiver-side physics-informed residual digital twin was developed for predictive fine tracking in inter-satellite optical links, combining nonlinear QPD/EKF synchronization, explicit nominal LOS propagation, and learned transition-mismatch correction while keeping delayed FSM actuation separate from physical LOS dynamics.
The four-way evaluation shows that feedback provides the main alignment improvement, while nominal constant-rate prediction offers little additional benefit at the 2-ms delay. PIR-Twin reduces mean RMS pointing error from 2.12 to 1.82~$\mu$rad and actuation-time prediction RMSE from about 1.95 to 1.62~$\mu$rad. It also maintains the lowest mean RMS error across the full 1--12-ms delay range. At 2~ms, its improvement over reactive tracking increases from 14.1\% under nominal LOS motion to 20.7\% and 23.7\% under $1.5\times$ and $2\times$ time-varying LOS-motion amplitudes, respectively, without residual-model retraining, whereas nominal prediction degrades rapidly as the prediction horizon increases. These results show that the predictive gain arises from correcting short-horizon model mismatch rather than from extrapolation alone. 
Overall, the results establish PIR-Twin as a robust receiver-side approach for delay-aware optical fine tracking, particularly when increasing actuator latency and LOS dynamics expose the limitations of nominal state extrapolation.

\section*{Acknowledgment}
This work was supported by the Qatar Research Development and Innovation Council (QRDI) under Grant No. NPRP14C-0909-210008 and by research funding from Hamad Bin Khalifa University under the Thematic Research Grant Program Cycle 3. The statements made herein are solely the responsibility of the authors. The content is solely the responsibility of the authors and does not necessarily represent the official views of QRDI.

\bibliographystyle{IEEEtran}
\bibliography{myref}

\end{document}